\documentclass{article}
\usepackage{babel}

\PassOptionsToPackage{numbers}{natbib}

\usepackage[final]{staix_2026}

\usepackage[utf8]{inputenc}
\usepackage[T1]{fontenc}
\usepackage{hyperref}
\usepackage{url}
\usepackage{booktabs}
\usepackage{amsfonts}
\usepackage{nicefrac}
\usepackage{microtype}
\usepackage{xcolor}
\usepackage{graphicx}
\graphicspath{{figures/}}
\usepackage{amsmath,amssymb,amsthm}
\usepackage{enumerate}
\usepackage{subcaption}
\usepackage{makecell}

\usepackage{listings}
\newcommand{\rev}[1]{{\color{black}#1}}
\newcommand{\E}{\mathbb{E}}
\newcommand{\Var}{\mathrm{Var}}

\newcommand{\tr}{\operatorname{tr}}
\newcommand{\diag}{\operatorname{diag}}

\theoremstyle{plain}
\newtheorem{theorem}{Theorem}[section]
\newtheorem{lemma}{Lemma}[section]
\newtheorem{proposition}{Proposition}[section]
\newtheorem{corollary}{Corollary}[section]
\newtheorem{assumption}{Assumption}[section]

\theoremstyle{definition}

\theoremstyle{remark}
\newtheorem{remark}{Remark}[section]

\title{Target-Aware Linear Regression \\ Under Distribution Shift}

\author{%
  Zhewen Hou\\
  Department of Statistics\\
  Columbia University\\
  New York, NY\\
  \texttt{zh2475@columbia.edu }
  \And
  Tian Zheng \\
  Department of Statistics\\
  Columbia University\\
  New York, NY\\
  \texttt{tian.zheng@columbia.edu}
}

\begin{document}

\maketitle

\begin{abstract}
Distribution shift between training and deployment is a pervasive challenge for modern AI systems. In many cases, the target marginals of covariates and response are known or specified through population-level observations, boundary conditions, properties of simulator configurations, or alignment-time distributional constraints. Such knowledge may provide valuable side information for regression estimation. We study this problem in the multivariate linear regression setting with a stable conditional mean $\mathbb{E}[Y\mid X]$ across source and target, and identify the hybrid-loss estimator, which jointly incorporates both target marginals, as \rev{a benchmark} target-aware estimator. Its direct computation, however, requires solving a coupled nonlinear optimization that is expensive at scale. Our main contribution is to develop and evaluate two computationally tractable alternatives: a constrained moment-matching estimator and a two-stage estimator that augments ordinary least squares with a calibration step. For all three estimators, we derive and compare closed-form asymptotic mean squared errors, yielding conditions under which the tractable alternatives match or closely approximate the hybrid benchmark, and regimes in which they do not. Monte Carlo experiments across three controlled shift regimes validate the theoretical results, investigate the accuracy–runtime tradeoffs among the three estimators, and translate into guidance on estimator choice. In particular, the two-stage estimator nearly matches the hybrid benchmark in the high signal-to-noise regime at essentially no additional cost, providing theoretical grounding for empirical observations in nonlinear settings.
\end{abstract}

\section{Introduction}

Modern AI models often face the challenge when the data they encounter during deployment differs from the data used during training.
When a model is trained on data drawn from a source distribution but evaluated under a different target distribution, the resulting mismatch can substantially degrade predictive performance.
In the literature, this problem has been addressed through importance weighting \citep{shimodaira2000improving}, domain adaptation \citep{ben2010theory}, and related techniques, typically assuming access to unlabeled samples from the target covariate distribution.

In many applications, however, the practitioner has access to richer side information about the target regime. Population level observations (e.g., income distributions) prescribe the marginal distribution of covariates in the deployment population.
Boundary conditions and conservation laws in physical systems constrain the marginal behavior of the response variable \citep{finzi2023user,zhong2023pi}.
Simulator configurations specify both input and output marginals for digital twins and climate projections \citep{gloege2021quantifying,kay2015community}.
Alignment-time distributional constraints in generative models fix the target output distribution by design.
\rev{In population studies, census data and administrative records provide precise marginal distributions of demographic and economic variables (e.g., age, income, education) for the target population, even when the training data come from a convenience sample or a different demographic.} \rev{In observational medical and epidemiological research, disease registries, vital statistics, and large-scale surveys furnish the population-level marginal distributions of clinical outcomes (e.g., blood pressure, BMI, mortality rates) and risk factors, against which study cohorts may be calibrated.} \rev{In survey methodology more broadly, known population totals from external sources are routinely used to calibrate survey weights so that weighted estimates match target marginals.}
In each of these settings, the \emph{target marginals} of covariates~$X$ and response~$Y$ are known or can be specified, and they provide valuable side information for regression estimation that goes beyond what unlabeled covariate data alone can offer.

In this paper, we study how to incorporate such target marginal information into regression estimation.
We work in the multivariate linear regression setting with Gaussian target marginals and a stable conditional mean $\mathbb{E}[Y \mid X]$ across source and target.
These assumptions are chosen to enable a rigorous and exact theoretical investigation, with closed-form asymptotic mean squared errors for every estimator for precise comparisons and sharp characterization of accuracy-complexity tradeoffs. The insights derived in this paper on
estimator ranking and the central role of signal-to-noise ratio confirm and provide theoretical grounding for empirical observations made in more complex, nonlinear settings.
In particular, \citet{hou2025calibrating} observe that post hoc calibration via the kernelized Stein discrepancy (KSD) yields the largest improvements when the base machine learning models are already reasonably accurate; our closed-form results explain precisely why this is so.

Within the proposed framework, we study three estimators.
The \emph{hybrid-loss estimator} jointly penalizes the empirical risk with a Wasserstein-type term that matches both target marginals simultaneously; it serves as \rev{a benchmark} target-aware estimator\rev{, uniformly dominating the tractable alternatives studied here, in terms of accuracy}.
Its computation, however, requires solving a coupled nonlinear optimization that is expensive at scale.
We therefore consider two computationally tractable alternatives.
The \emph{constrained moment-matching estimator} enforces the target marginal constraints exactly.
The \emph{two-stage estimator} first fits ordinary least squares and then calibrates the predictions to match the target marginals in a post-processing step.

Our main \textbf{contributions} are as follows.
\begin{enumerate}
\item We formulate the target-aware regression problem in a multivariate linear model and develop three estimators: hybrid, constrained moment-matching, and two-stage, all of which incorporate target marginal information in structurally different ways.
\item For all three estimators, we derive the asymptotic distribution, mean squared error, and prediction error, making the accuracy-complexity tradeoffs transparent.
\item We identify precise conditions under which the tractable alternatives match or closely approximate the hybrid benchmark, as well as regimes in which they do not. A key finding is that the two-stage estimator, which requires only a scalar rescaling of OLS, nearly matches the hybrid estimator in the high signal-to-noise regime, which is the practically relevant setting when the base model is already powerful.
\item Monte Carlo experiments across three controlled shift regimes validate the theoretical results, demonstrate accuracy-runtime tradeoffs, and provide guidance on estimator choice in practice.
\end{enumerate}

\section{Related Work}

{\bf Covariate shift.}
When the covariate distribution changes from source to target but the conditional $Y\mid X$ remains stable, importance weighting \citep{shimodaira2000improving} reweights the training loss by the density ratio $p_t(x)/p_s(x)$.
Estimating this ratio in high dimensions is itself challenging; approaches include low-dimensional projections \citep{stojanov2019low}, adaptive kernel methods \citep{zellinger2023adaptive}, and nearest-neighbor sampling \citep{portier2024nearest}.
These methods use knowledge of the target covariate distribution~$\mathcal{P}_{t,x}$ but do not incorporate the target response marginal~$\mathcal{P}_{t,y}$.
\rev{However, importance-weighting estimators do not improve upon OLS and are less favorable in the regimes we analyze. This contrasts with the estimators studied in this paper, whose main advantage comes from incorporating the target response marginal~$\mathcal{P}_{t,y}$ in addition to target covariate information. Thus, importance weighting and target-marginal estimation should be viewed as complementary: the former corrects covariate mismatch through reweighting, while the latter leverages response-marginal information to reduce estimation variance.}

{\bf Distributional learning.}
Wasserstein Distributional Learning (WDL) \citep{tang2023wasserstein} proposes a majorization-minimization algorithm for fitting semi-parametric conditional Gaussian mixture models for conditional density estimation, using the Wasserstein distance $W_2$ as a metric on density outcomes. It highlights the practical feasibility of Wasserstein-based distributional matching.

{\bf Post-hoc distribution calibration.} In the nonlinear setting, \citet{hou2025calibrating} propose a post hoc calibration method based on the kernelized Stein discrepancy (KSD) \citep{liu2016kernelized} that adjusts predictions from an arbitrary base model to match a known target distribution. A normalization step rescales model-generated predictions to match target moments before KSD refinement. This is a nonlinear analog of the two-stage estimator studied here. \citet{hou2025calibrating} provided intuition on the MSE reduction of the KSD gradient step via Stein's Lemma, directly connecting to the James-Stein mechanism.

\rev{{\bf Constrained and auxiliary-information estimation.}}
\rev{The moment-matching estimator can be viewed as a constrained least-squares estimator \citep{judge1980theory}.}
\rev{Many methods have been developed to use known population information to improve efficiency, including survey calibration \citep{devaud2019deville,deville1992calibration,sarndal2007calibration}, the generalized method of moments (GMM) with auxiliary moments \citep{chen2008semiparametric,hansen1982large,imbens1997one}, and empirical likelihood with auxiliary summary information \citep{gao2023noniterative,han2019empirical,qin1994empirical}.}
\rev{Recent semi-supervised and calibrated-estimation work also exploits unlabeled covariates, calibration losses, or balancing constraints to reduce variance \citep{bruns2025augmented,chakrabortty2018efficient,hirshberg2021augmented,tan2020regularized}.}
\rev{In our setting, the auxiliary information is the target response marginal.}
\rev{The moment-matching estimator enforces the target mean and variance exactly, while the hybrid estimator relaxes these constraints through a penalty.}
\rev{The main distinction is that the constraints are target-distributional rather than source-sample moment restrictions or known covariate totals.}


\rev{{\bf Shrinkage estimation and external information.}}
\rev{
The two-stage estimator bears a structural resemblance to James--Stein shrinkage
\citep{efron1973stein,stein1956inadmissibility}: both use external information to suppress unnecessary sampling variation in selected directions.
\citet{han2024improving} study James-Stein-type estimators that incorporate external population information into linear regression.
In our setting, however, the target marginal constraints are valid population constraints satisfied by the true parameter, so the leading-order gain is an efficiency gain from auxiliary information rather than a leading-order bias effect.
}

\section{Methods}

We formalize the target-aware regression problem in a multivariate linear model with known target marginals. After establishing the setup and notation, we present the OLS baseline and three target-aware estimators. For every estimator we state the asymptotic distribution, mean squared error, and prediction error under the target distribution.

\subsection{Setup and notation}
We consider the multivariate linear regression model.

\begin{assumption}
\label{assump:multi_linear_model}
Let $X\in\mathbb{R}^d$ be a $d$-dimensional covariate vector. The response variable satisfies
\[
Y = \beta_0 + X^\top \beta + \varepsilon,
\qquad
\varepsilon \sim \mathcal{N}(0,\sigma_{\varepsilon}^2),
\]
where $\beta\in\mathbb{R}^d$ and $X \perp \varepsilon$.
\end{assumption}

For simplicity, write the augmented regressor as $\tilde{X} := (1, X^\top)^\top \in \mathbb{R}^{d+1}$, $\theta := (\beta_0,\beta^\top)^\top \in \mathbb{R}^{d+1}$.

\begin{assumption}
\label{assump:multi_training_sample}
We observe i.i.d.\ data $\{(X_i,Y_i)\}_{i=1}^n$ generated (from the source distribution, subscript $s$) as
\[
Y_i = \beta_0 + X_i^\top \beta + \varepsilon_i,
\qquad i=1,\dots,n.
\]
Moreover, $X_i \sim \mathcal{P}_{s,x}$, $\mathbb{E}_{\mathcal{P}_{s,x}}[X] = \mu_{s,x}\in\mathbb{R}^d$, $\operatorname{Cov}_{\mathcal{P}_{s,x}}(X) = \Sigma_{s,x}\in\mathbb{R}^{d\times d}$,
where $\Sigma_{s,x}$ is positive definite.
The distribution $\mathcal{P}_{s,x}$ may be unknown.
\end{assumption}

\begin{assumption}
\label{assump:multi_fourth_moment}
The training covariate satisfies $\mathbb{E}_{\mathcal{P}_{s,x}}\|X\|^4<\infty$.
\end{assumption}
\begin{assumption}
\label{assump:multi_subgaussian}
The training covariate is sub-Gaussian under $\mathcal{P}_{s,x}$.
\end{assumption}

Let $\boldsymbol{Y}=(Y_1,\dots,Y_n)^\top$ and $\boldsymbol{\tilde X}=(\tilde X_1,\dots,\tilde X_n)^\top\in\mathbb{R}^{n\times(d+1)}$. Define the source moment matrix $Q_s:=\E_{\mathcal{P}_{s,x}}(\tilde X \tilde X^\top)$.

A key structural assumption is that the conditional mean $\mathbb{E}[Y\mid X]=\beta_0+X^\top\beta$ is \emph{stable} across source and target: the regression coefficients $(\beta_0,\beta)$ and the noise variance $\sigma_\varepsilon^2$ do not change, while the marginal distribution of $X$ may differ.

In addition to the regression structure, we assume access to target marginals describing the marginal behavior of the model variables under the deployment distribution.
\begin{assumption}
\label{assump:multi_knowledge_distribution}
The target marginals of $(X,Y)$ are known (subscript $k$) and specified as
\[
X \sim \mathcal{N}(\mu_{k,x},\Sigma_{k,x}),
\qquad
Y \sim \mathcal{N}(\mu_{k,y},\sigma_{k,y}^2),
\]
where $\mu_{k,x}\in\mathbb{R}^d$ and $\Sigma_{k,x}\in\mathbb{R}^{d\times d}$ is positive definite.
\end{assumption}

Write $\tilde{\mu}_{k,x} = (1,\mu_{k,x}^\top)^\top$ and $\tilde{\Sigma}_{k,x}=\operatorname{Var}(\tilde X)=\diag(0,\Sigma_{k,x})$.
Under Assumptions~\ref{assump:multi_linear_model} and \ref{assump:multi_knowledge_distribution}, $\mathbb{E}[Y] = \beta_0 + \mu_{k,x}^\top \beta = \mu_{k,y}$ and $\operatorname{Var}(Y)= \beta^\top \Sigma_{k,x}\beta + \sigma_\varepsilon^2 = \sigma_{k,y}^2$. The source and target marginals need not coincide.

\smallskip
\begin{remark}
\label{rem:multi_normalization}
By applying a linear transformation to $X$, we may assume without loss of generality that
$\mu_{k,x}=0$ and $\Sigma_{k,x}=I_d$.
This changes $\theta$, $\mu_{s,x}$, and $\Sigma_{s,x}$, but preserves the linear relationship, $\sigma_\varepsilon$, $\mu_{k,y}$, and $\sigma_{k,y}$.
\end{remark}

\subsection{OLS baseline}
The ordinary least squares (OLS) estimator is
$\hat\theta^{\mathrm{OLS}}
=
(\boldsymbol{\tilde X}^{\top}\boldsymbol{\tilde X})^{-1}
\boldsymbol{\tilde X}^{\top}\boldsymbol{Y}$,
with noise variance estimator
$\hat\sigma_\varepsilon^2
=
\|\boldsymbol{Y}-\boldsymbol{\tilde X}\hat\theta^{\mathrm{OLS}}\|_2^2/(n-d-1)$.
By standard arguments, $\sqrt{n}(\hat\theta^{\mathrm{OLS}}-\theta)
\Rightarrow
\mathcal{N}(0,\sigma_\varepsilon^2 Q_s^{-1})$,
giving 
\[
\mathbb{E}\|\hat\beta^{\mathrm{OLS}}-\beta\|^2
=
(\sigma_\varepsilon^2/n)\operatorname{tr}(\Sigma_{s,x}^{-1})
+o(n^{-1}).
\]
OLS ignores the target marginals, and its prediction error under the target distribution,
\begin{equation}
\mathbb{E}\!\left[\left(Y-\tilde{X}^\top\hat\theta^{\mathrm{OLS}}\right)^2\right]
=
\sigma_\varepsilon^2
+
\frac{\sigma_\varepsilon^2}{n}
\left[
1+\operatorname{tr}\!\left(\Sigma_{k,x}\Sigma_{s,x}^{-1}\right)
+
(\mu_{k,x}-\mu_{s,x})^\top \Sigma_{s,x}^{-1}(\mu_{k,x}-\mu_{s,x})
\right]
+
o(n^{-1}),
\label{eq:OLS_prediction_error}
\end{equation}
grows with the source--target mismatch through both the covariance ratio and the squared Mahalanobis distance of the mean shift.

\subsection{Hybrid-loss estimator}
The hybrid estimator combines the empirical risk with a penalty that matches the model-implied output distribution to the target marginals:
\begin{equation}
L_{\mathrm{H}}(\theta;\omega)
=
\frac{1}{n}\,
\| \boldsymbol{Y} - \boldsymbol{\tilde X} \theta \|_2^2
+
\omega\left(\tilde{\mu}_{k,x}^\top\theta-\mu_{k,y}\right)^2
+
\omega
\left(
\sqrt{
\theta^\top\tilde\Sigma_{k,x}\theta
+
\frac{1}{n}\,
\| \boldsymbol{Y} - \boldsymbol{\tilde X} \theta \|_2^2
}
-\sigma_{k,y}
\right)^2,
\label{eq:hybrid_loss_multi}
\end{equation}
where $\omega\ge 0$ controls the weight of the target constraints. The second term enforces mean matching, and the third enforces variance matching via the squared Wasserstein distance between two one-dimensional Gaussians. We define
$\hat\theta^{\mathrm{H}}(\omega)\in\arg\min_{\theta} L_{\mathrm{H}}(\theta;\omega)$.
Since the loss is nonlinear in $\theta$, the minimizer is computed numerically by gradient descent initialized at $\hat\theta^{\mathrm{OLS}}$; the explicit gradient is given in Section~\ref{app:hybrid_detail}.

Let $\tilde{v}_{\sigma\beta}=\tilde\Sigma_{k,x}\theta$ and define
\[
Q(\omega)
:=
Q_s
+
\omega\!\left(
\tilde\mu_{k,x}\tilde\mu_{k,x}^\top
+
\frac{1}{\sigma_{k,y}^2}\tilde v_{\sigma\beta}\tilde v_{\sigma\beta}^\top
\right),
\qquad
\Omega(\omega)
:=
Q_s
+
\frac{\omega^2\sigma_\varepsilon^2}{2\sigma_{k,y}^4}
\tilde v_{\sigma\beta}\tilde v_{\sigma\beta}^\top.
\]

All proofs are deferred to Section~\ref{app:proofs} of the Supplement; detailed derivations of the estimators, including the constraint geometry, the gradient and quartic characterization of the optimal weight, and the prediction error formulas, are given in Section~\ref{app:derivations}.

\smallskip
\begin{theorem}
\label{thm:hybrid_multi_asymp}
Under Assumptions~\ref{assump:multi_linear_model}--\ref{assump:multi_knowledge_distribution}, for any $\omega\ge0$, as $n\to\infty$,
\[
\sqrt{n}\left(\hat\theta^{\mathrm{H}}(\omega)-\theta\right)
\Rightarrow
\mathcal{N}\!\left(0,\,\Sigma_{\mathrm H}(\omega)\right),
\qquad
\Sigma_{\mathrm H}(\omega)
=
\sigma_\varepsilon^2\,
Q(\omega)^{-1}\Omega(\omega)Q(\omega)^{-1}.
\]
\end{theorem}

\begin{corollary}
\label{cor:hybrid_multi_beta_mse}
Under the same assumptions, the asymptotic mean squared error is
\begin{equation}
\mathbb{E}\!\left[\left\|\hat\beta^{\mathrm H}(\omega)-\beta\right\|^2\right]
=
\frac{\sigma_\varepsilon^2}{n}\,
V_{\mathrm H}(\omega)
+
o(n^{-1}),
\label{eqn_beta_H_MSE}
\end{equation}
where $V_{\mathrm H}(\omega)
:=
\operatorname{tr}\!\left(
P_\beta Q(\omega)^{-1}\Omega(\omega)Q(\omega)^{-1}P_\beta^\top
\right)$ with $P_\beta:=(0_{d\times 1},I_d)$.

Under the normalization of Remark~\ref{rem:multi_normalization}, define
$M_{\mathrm H}(\omega)
:=
\Sigma_{s,x}
+
\frac{\omega}{1+\omega}\mu_{s,x}\mu_{s,x}^\top
+
\frac{\omega}{\sigma_{k,y}^2}\beta\beta^\top$.
Then eq.\ (\ref{eqn_beta_H_MSE}) becomes
\begin{equation}
    \begin{aligned}
        \mathbb{E}\!\left[\left\|\hat\beta^{\mathrm H}(\omega)-\beta\right\|^2\right]
&=&
\frac{\sigma_\varepsilon^2}{n}
\left[
\operatorname{tr}\!\left(M_{\mathrm H}(\omega)^{-1}\right)
-
\frac{\omega}{(1+\omega)^2}
\mu_{s,x}^\top M_{\mathrm H}(\omega)^{-2}\mu_{s,x} \right. \\
&&
\left. -
\left(
\frac{\omega}{\sigma_{k,y}^2} 
-
\frac{\omega^2\sigma_\varepsilon^2}{2\sigma_{k,y}^4}
\right)
\beta^\top M_{\mathrm H}(\omega)^{-2}\beta
\right]
+
o(n^{-1}).
    \end{aligned}
    \label{eq:hybrid_mse_normalized}
\end{equation}

\end{corollary}

\begin{remark}
Here, the matrix $Q(\omega)$ augments the source moment $Q_s$ with two rank-one terms from the mean and variance constraints. The matrix $\Omega(\omega)$ differs from $Q(\omega)$ because of the variance constraint. 
The three terms in eq.\ (\ref{eq:hybrid_mse_normalized}) have clear roles: the first is the baseline variance under the augmented geometry. The second is the reduction from the mean constraint, weighted by the mean shift $\mu_{s,x}$; and the third is the ``reduction'' from the variance constraint, which is positive (beneficial) when $\omega\sigma_\varepsilon^2/(2\sigma_{k,y}^2)<1$ and changes sign at large noise levels.
   
\end{remark}

\paragraph{Optimal $\omega^\star$.}
The asymptotically optimal tuning parameter is $\omega^\star\in\arg\min_{\omega\ge 0}V_{\mathrm H}(\omega)$.
The function $V_{\mathrm H}(\omega)$ is a rational function of $\omega$ (Proposition~\ref{prop:VH_rational}) whose derivative admits a quartic equation. In practice, one collects its nonnegative real roots and compares $V_{\mathrm H}$ at those roots and at $\omega=0$.

\subsection{Constrained moment-matching (MM) estimator}
Rather than penalizing the Wasserstein distance, the moment-matching estimator enforces the target marginal constraints exactly; equivalently, it replaces the soft penalty in \eqref{eq:hybrid_loss_multi} by a hard constraint, so that the hybrid loss with $\omega\to\infty$ recovers the moment-matching objective. Define the loss
\begin{equation}
L_{\mathrm{MM}}(\theta)
=
\left(\tilde{\mu}_{k,x}^\top\theta-\mu_{k,y}\right)^2
+
\left(
\sqrt{
\theta^\top\tilde\Sigma_{k,x}\theta+\frac{1}{n}\|\boldsymbol Y-\boldsymbol{\tilde X}\theta\|_2^2
}
-\sigma_{k,y}
\right)^2.
\label{eq:wdl_loss_multi}
\end{equation}
Let $\hat\Theta^{\mathrm{MM}}=\arg\min_\theta L_{\mathrm{MM}}(\theta)$. The mean constraint fixes $\beta_0=\mu_{k,y}-\mu_{k,x}^\top\beta$, and the variance constraint reduces to a quadratic in~$\beta$ (Lemma~\ref{lem:wdl_zero_loss}). When $d>1$ and the zero-loss set is non-singleton, we select the element with smallest residual sum of squares, which is characterized by a KKT system with a single Lagrange multiplier (Proposition~\ref{prop:wdl_kkt}):
\[
\hat\theta^{\mathrm{MM}}
=
\arg\min_{\theta\in\hat\Theta^{\mathrm{MM}}}
\sum_{i=1}^n (\check Y_i-\check X_i^\top\beta)^2,
\qquad
\check X_i:=X_i-\mu_{k,x},\ \check Y_i:=Y_i-\mu_{k,y}.
\]

Write $Q_{s\mid k}:=\Sigma_{s,x}+(\mu_{s,x}-\mu_{k,x})(\mu_{s,x}-\mu_{k,x})^\top$, $v_{\sigma\beta}:=\Sigma_{k,x}\beta$, $\kappa:=v_{\sigma\beta}^\top Q_{s\mid k}^{-1}v_{\sigma\beta}$, $\chi:=v_{\sigma\beta}^\top Q_{s\mid k}^{-1}\Sigma_{k,x}Q_{s\mid k}^{-1}v_{\sigma\beta}$, $J_{\mu_k}:=\bigl(\begin{smallmatrix}\mu_{k,x}^\top\\ -I_d\end{smallmatrix}\bigr)$, and $\Omega
:=
Q_{s\mid k}
-
\frac{v_{\sigma\beta}v_{\sigma\beta}^\top}{\kappa}
+
\frac{\sigma_\varepsilon^2}{2}
\frac{v_{\sigma\beta}v_{\sigma\beta}^\top}{\kappa^2}$.

\begin{theorem}
\label{thm:wdl_multi_asymp}
Under Assumptions~\ref{assump:multi_linear_model}--\ref{assump:multi_knowledge_distribution}, if $\beta\neq 0$, as $n\to\infty$,
\[
\sqrt{n}\left(\hat\theta^{\mathrm{MM}}-\theta\right)
\Rightarrow
\mathcal{N}\!\left(0,\,\Sigma_{\mathrm{MM}}\right),
\qquad
\Sigma_{\mathrm{MM}}
=
\sigma_\varepsilon^2\,
J_{\mu_k}Q_{s\mid k}^{-1}\Omega\, Q_{s\mid k}^{-1}J_{\mu_k}^\top.
\]
\end{theorem}

\begin{corollary}
\label{cor:wdl_multi_beta_mse}
Under the same assumptions, 
\[
\mathbb{E}\!\left[\left\|\hat\beta^{\mathrm{MM}}-\beta\right\|^2\right]
=
\frac{\sigma_\varepsilon^2}{n}\,
\operatorname{tr}\!\left(
Q_{s\mid k}^{-1}\Omega\, Q_{s\mid k}^{-1}
\right)
+
o(n^{-1}).
\]
Under the normalization of Remark~\ref{rem:multi_normalization}, this becomes
\begin{equation}
\frac{\sigma_\varepsilon^2}{n}
\left[
\operatorname{tr}\!\left(Q_{s\mid k}^{-1}\right)
-
\frac{\beta^\top Q_{s\mid k}^{-2}\beta}{\beta^\top Q_{s\mid k}^{-1}\beta}
+
\frac{\sigma_\varepsilon^2}{2}
\frac{\beta^\top Q_{s\mid k}^{-2}\beta}{\left(\beta^\top Q_{s\mid k}^{-1}\beta\right)^2}
\right]
+
o(n^{-1}).
\label{eq:wdl_mse}
\end{equation}
The moment-matching estimator improves over OLS when $\sigma_\varepsilon^2\le 2\beta^\top Q_{s\mid k}^{-1}\beta$.
\end{corollary}

\begin{remark}
The quantity $\beta^\top Q_{s\mid k}^{-1}\beta$ measures the effective signal strength under the geometry induced by the training distribution, while $\sigma_\varepsilon^2$ is the noise level. When noise is not too large relative to this effective signal, the target marginal constraints shrink the feasible set of $\beta$ and suppress unnecessary sampling variation, yielding a smaller asymptotic MSE than OLS.    
\end{remark}

\subsection{Two-stage calibration estimator}
The two-stage estimator first computes OLS and then calibrates the predictions to match the target marginals. We work under the normalization of Remark~\ref{rem:multi_normalization}. Suppose an independent test sample $\{X_j^{\mathrm{test}}\}_{j=1}^m$ is drawn from $\mathcal{N}(0,I_d)$ with empirical moments matching the target exactly.

\emph{Stage 1.} Compute $\hat\theta^{\mathrm{OLS}}$ and $\hat\sigma_\varepsilon^2$ from the training data.

\emph{Stage 2.} Form raw predictions $\hat Y_j=\hat\beta_0^{\mathrm{OLS}}+(X_j^{\mathrm{test}})^\top\hat\beta^{\mathrm{OLS}}$ and rescale them to match the target mean $\mu_{k,y}$ and target prediction variance $(\sigma_{k,y}^2-\hat\sigma_\varepsilon^2)_+$. This yields calibrated predictions $\tilde Y_j = a + b\,\hat Y_j$ with
\[
a = \mu_{k,y} - b\cdot\bar{\hat Y},
\qquad
b = \sqrt{\frac{(\sigma_{k,y}^2-\hat\sigma_\varepsilon^2)_+}{\|\hat\beta^{\mathrm{OLS}}\|^2}}.
\]
The implied coefficient estimator is $\hat\beta^{\mathrm{cali}} = b\cdot\hat\beta^{\mathrm{OLS}}$. The target mean constraint determines the intercept exactly ($\hat\beta_0^{\mathrm{cali}}=\mu_{k,y}-\mu_{k,x}^\top\hat\beta^{\mathrm{cali}}$), so the MSE below accounts only for the slope.

\begin{theorem}
\label{thm:cali_multi}
Under Assumptions~\ref{assump:multi_linear_model}--\ref{assump:multi_knowledge_distribution} and the normalization of Remark~\ref{rem:multi_normalization}, with $\beta\neq 0$,
\begin{equation}
\mathbb{E}\!\left[\left\|\hat\beta^{\mathrm{cali}}-\beta\right\|^2\right]
=
\frac{\sigma_\varepsilon^2}{n}
\left[
\operatorname{tr}\!\left(\Sigma_{s,x}^{-1}\right)
-
\frac{\beta^\top \Sigma_{s,x}^{-1}\beta}{\|\beta\|^2}
+
\frac{\sigma_\varepsilon^2}{2\|\beta\|^2}
\right]
+
o(n^{-1}).
\label{eq:cali_mse}
\end{equation}
The prediction error satisfies
\begin{equation}
\mathbb{E}\!\left[\frac{1}{m}\sum_{j=1}^m(\tilde Y_j-Y_j^{\mathrm{test}})^2\right]
=
\sigma_\varepsilon^2
+
\frac{\sigma_\varepsilon^2}{n}
\left[
\operatorname{tr}\!\left(\Sigma_{s,x}^{-1}\right)
-
\frac{\beta^\top \Sigma_{s,x}^{-1}\beta}{\|\beta\|^2}
+
\frac{\sigma_\varepsilon^2}{2\|\beta\|^2}
\right]
+
o(n^{-1}).
\label{eq:cali_prediction}
\end{equation}
\end{theorem}

\begin{remark}
  The two-stage rescaling reduces the variance along the signal direction $\beta$ at the cost of introducing a noise-dependent correction $\sigma_\varepsilon^2/(2\|\beta\|^2)$.
\end{remark}

\begin{remark}[The $\beta=0$ boundary]
\label{rem:beta_zero}
\rev{Theorems~\ref{thm:wdl_multi_asymp} and \ref{thm:cali_multi} require $\beta\neq 0$.}
\rev{When $\beta=0$, the response $Y=\beta_0+\varepsilon$ is pure noise and $\sigma_{k,y}^2=\sigma_\varepsilon^2$.}
\rev{The target variance constraint carries no directional information about $\beta$, and the rescaling factor $b$ is ill-defined (division by $\|\hat\beta^{\mathrm{OLS}}\|^2\to 0$).}
\rev{In this degenerate regime, OLS is already efficient for estimating the slope, so the target marginals offer no additional gain for inference on $\beta$ (although the mean constraint still suffices to identify $\beta_0$).}
\rev{When $\|\beta\|$ is small but nonzero, the correction terms $\sigma_\varepsilon^2/(2\|\beta\|^2)$ and $\sigma_\varepsilon^2/(2\kappa)$ grow, and the target-aware estimators may perform worse than OLS; the improvement condition $\sigma_\varepsilon^2 < 2\beta^\top Q_{s\mid k}^{-1}\beta$ correctly excludes this regime.}
\end{remark}

\smallskip
\textbf{Computational complexity of estimators.}

As shown in Table~\ref{tab:comparison} , OLS, two-stage, and moment-matching (MM) all have essentially the same computational cost as OLS (matrix inversion plus only a scalar rescaling or a simple root-finding step), whereas the hybrid estimator requires tolerance-dependent iterative optimization.


\begin{table}[h]
\centering
\caption{Computational costs of different estimators.}
\label{tab:comparison}
\begin{tabular}{lccc}
\toprule
Estimator & Computation & Relative cost \\
\midrule
OLS & matrix inversion & $1\times$ \\
Two-stage (cali)  & OLS $+$ scalar rescaling & ${\approx}\,1\times$ \\
Moment-matching (MM)  & OLS $+$ root-finding & ${\approx}\,1\times$ \\
Hybrid & iterative optimization & tolerance-dependent \\
\bottomrule
\end{tabular}
\end{table}

\section{Numerical Experiments}
\label{sec:experiments}

\textbf{Setup.}
We perform experiments in dimension $d=3$ under the multivariate linear model with intercept $\beta_0=1$ and default coefficients $\beta=(0.970,-1.268,0.671)^\top$. The target distribution is normalized: $\mu_{k,x}=0$, $\Sigma_{k,x}=I_3$. Default source parameters are $\mu_{s,x}=0$, $\Sigma_{s,x}=I_3$.

For each configuration, we generate $L=10^6$ independent Monte Carlo repetitions with training sample size $n=1000$ and test sample size $m=100$ (with test covariates exactly matching target moments).
\rev{We compare four estimators: OLS, moment-matching (MM), hybrid, and two-stage (cali). For the hybrid estimator, we compute a plug-in $\hat\omega$ by replacing the population quantities in the characterization of $\omega^\star$ in Proposition~\ref{prop:VH_rational} with sample estimates. We then minimize $L_{\mathrm H}(\theta;\hat\omega)$ from the OLS initialization using L-BFGS-B with maximum iteration count $1000$, $\texttt{ftol}=10^{-12}$, and $\texttt{gtol}=10^{-8}$.} 
We report two metrics: the scaled coefficient error $n\|\hat\beta-\beta\|_2^2/\sigma_\varepsilon^2$ and the scaled excess prediction error $n(\mathrm{MSE}_{\mathrm{test}}-\sigma_\varepsilon^2)/\sigma_\varepsilon^2$.

We consider three controlled regimes, each varying one aspect of the source--target relationship while holding the others fixed.

\smallskip
\textbf{Noise level.}
We vary $\sigma_\varepsilon^2\in\{1,2,\ldots,20\}$ with $\mu_{s,x}=0$, $\Sigma_{s,x}=I_3$. Under the isotropic setting, the theoretical scaled MSEs are $3$ for OLS, $2+\sigma_\varepsilon^2/(2\|\beta\|^2)$ for moment-matching and two-stage, and $2+\sigma_\varepsilon^2/(2\|\beta\|^2+\sigma_\varepsilon^2)$ for the hybrid. Moment-matching and two-stage coincide and beat OLS only when $\sigma_\varepsilon^2<2\|\beta\|^2\approx 6$.
\begin{figure}[h]
    \centering
    \includegraphics[width=0.5\textwidth]{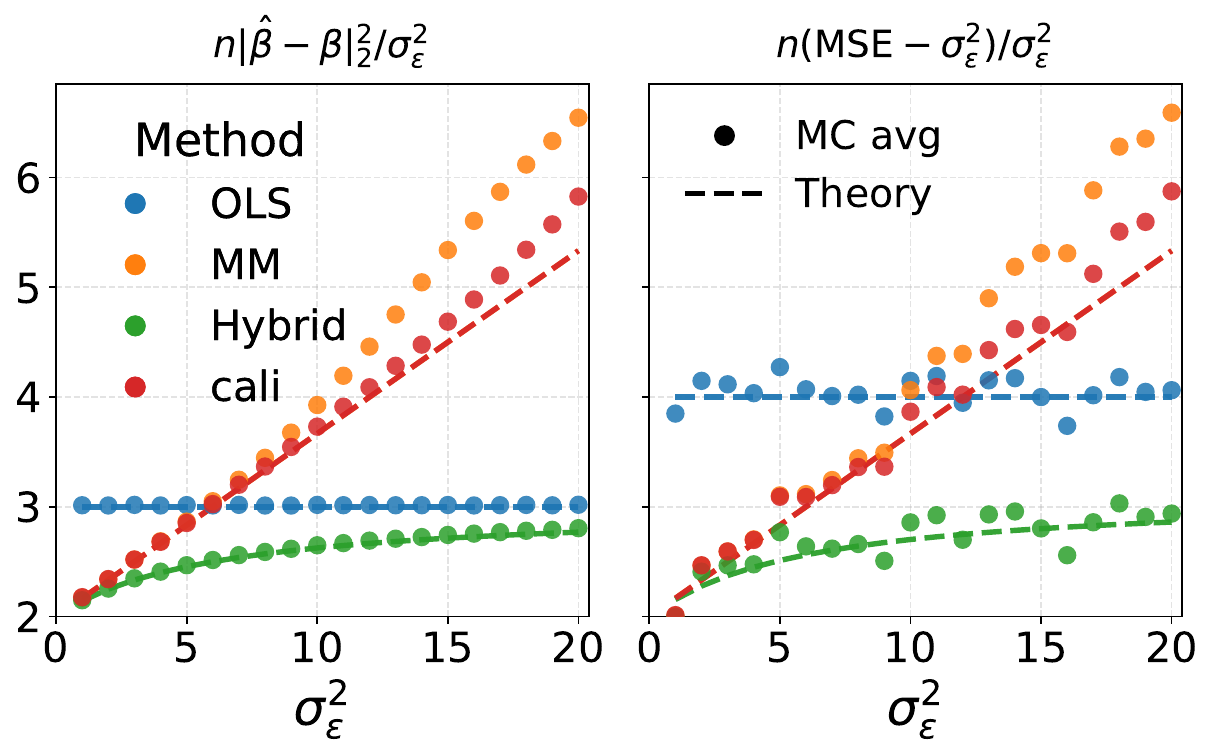}
    \caption{\textbf{Noise level.} Scaled coefficient error (left) and scaled excess prediction error (right) vs.\ $\sigma_\varepsilon^2$. Points: Monte Carlo averages ($L=10^6$, $n=1000$); dashed lines: theoretical values. \rev{Monte Carlo standard errors are on the order of $10^{-3}$ and are visually negligible at this scale.}}
    \label{fig:toy_sigma_eps}
\end{figure}
Figure~\ref{fig:toy_sigma_eps} confirms the theoretical results: the OLS curve is flat at $3$, the moment-matching and two-stage curves rise linearly, and the hybrid stays below them throughout. Target marginals provide the largest gains in the high signal-to-noise regime.

\smallskip
\textbf{Covariance geometry.}
We vary $\Sigma_{s,x}$ subject to $\tr(\Sigma_{s,x}^{-1})=3$ and $\beta^\top\Sigma_{s,x}^{-1}\beta\in\{0.5,1.0,\ldots,8.5\}$, displaying results at $\sigma_\varepsilon^2=2$ where the geometry effect is largest. The general MSE formulas \eqref{eq:wdl_mse} and \eqref{eq:cali_mse} now depend on the alignment between $\Sigma_{s,x}$ and $\beta$, and the two estimators are no longer identical.
\begin{figure}[h]
    \centering
    \includegraphics[width=0.75\textwidth]{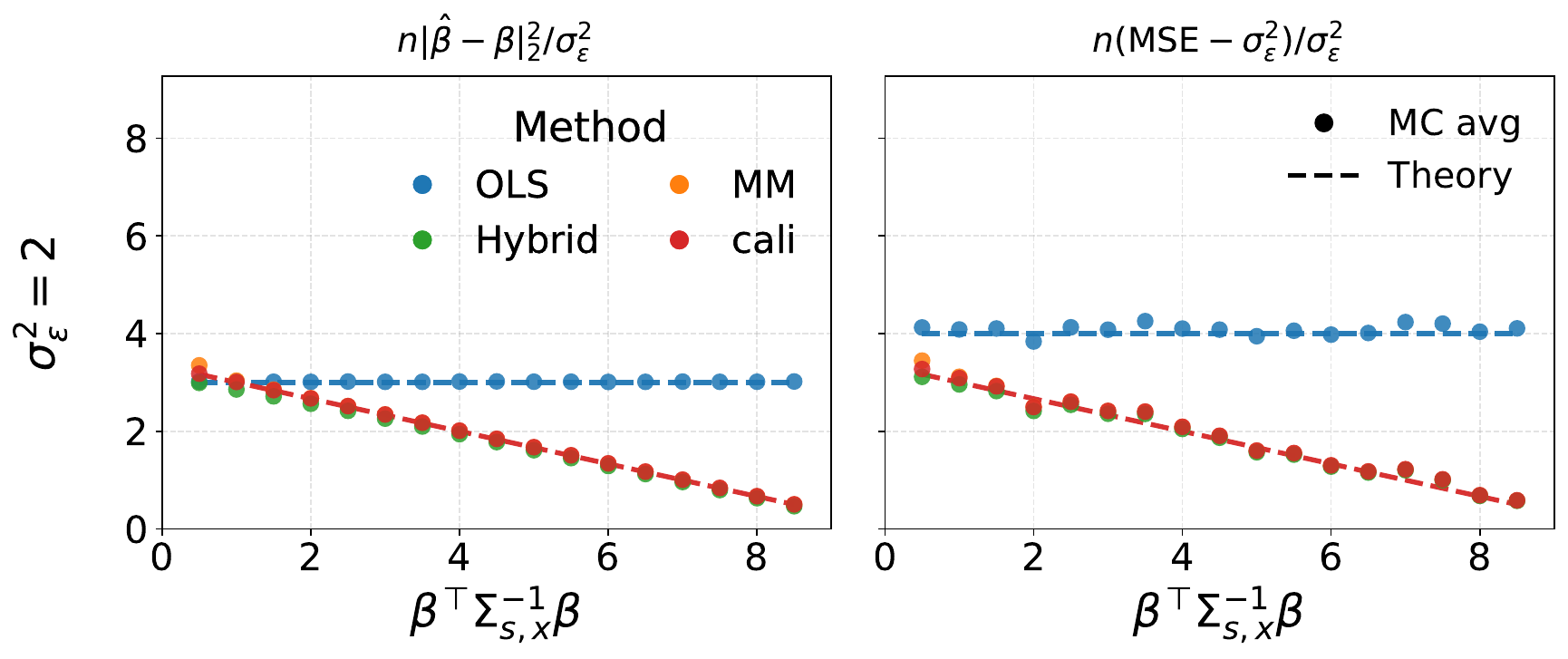}
    \caption{\textbf{Covariance geometry} ($\sigma_\varepsilon^2=2$). Scaled coefficient error (left) and excess prediction error (right) vs.\ $\beta^\top\Sigma_{s,x}^{-1}\beta$. \rev{Monte Carlo standard errors are on the order of $10^{-3}$ and are visually negligible at this scale.}}
    \label{fig:toy_sigma_sx}
\end{figure}
Figure~\ref{fig:toy_sigma_sx} shows $\sigma_\varepsilon^2=2$, where the geometry effect is most pronounced. Notably, at low values of $\beta^\top\Sigma_{s,x}^{-1}\beta$, moment-matching is worse than OLS: since $\mu_{s,x}=0$ here, the improvement condition $\sigma_\varepsilon^2<2\beta^\top Q_{s\mid k}^{-1}\beta$ (Corollary~\ref{cor:wdl_multi_beta_mse}) reduces to $\beta^\top\Sigma_{s,x}^{-1}\beta>\sigma_\varepsilon^2/2=1$, and the figure confirms that moment-matching crosses OLS near this threshold. As $\beta^\top\Sigma_{s,x}^{-1}\beta$ increases, both moment-matching and two-stage improve, but moment-matching improves faster because it exploits the full geometry of $Q_{s\mid k}$, whereas two-stage projects onto $\beta$. The hybrid remains uniformly best. The full results across multiple noise levels (Figure~\ref{fig:supp_sigma_sx} in the Supplement) reveal that at low noise the three target-aware estimators are nearly indistinguishable and they separate as the noise increases.

\smallskip
\textbf{Mean mismatch.}
We fix $\|\mu_{s,x}\|^2=3$, $\Sigma_{s,x}=I_3$, and vary $\rho=\|P_{\beta^\perp}\mu_{s,x}\|^2/\|\mu_{s,x}\|^2\in\{0,0.1,\ldots,1.0\}$, where $P_{\beta^\perp}=I_d-\beta\beta^\top/\|\beta\|^2$. Since $\Sigma_{s,x}=I_3$, the two-stage MSE \eqref{eq:cali_mse} is flat in $\rho$, whereas the moment-matching MSE depends on the direction of the mean shift through $Q_{s\mid k}$.

\begin{figure}[h]
    \centering
    \includegraphics[width=0.75\textwidth]{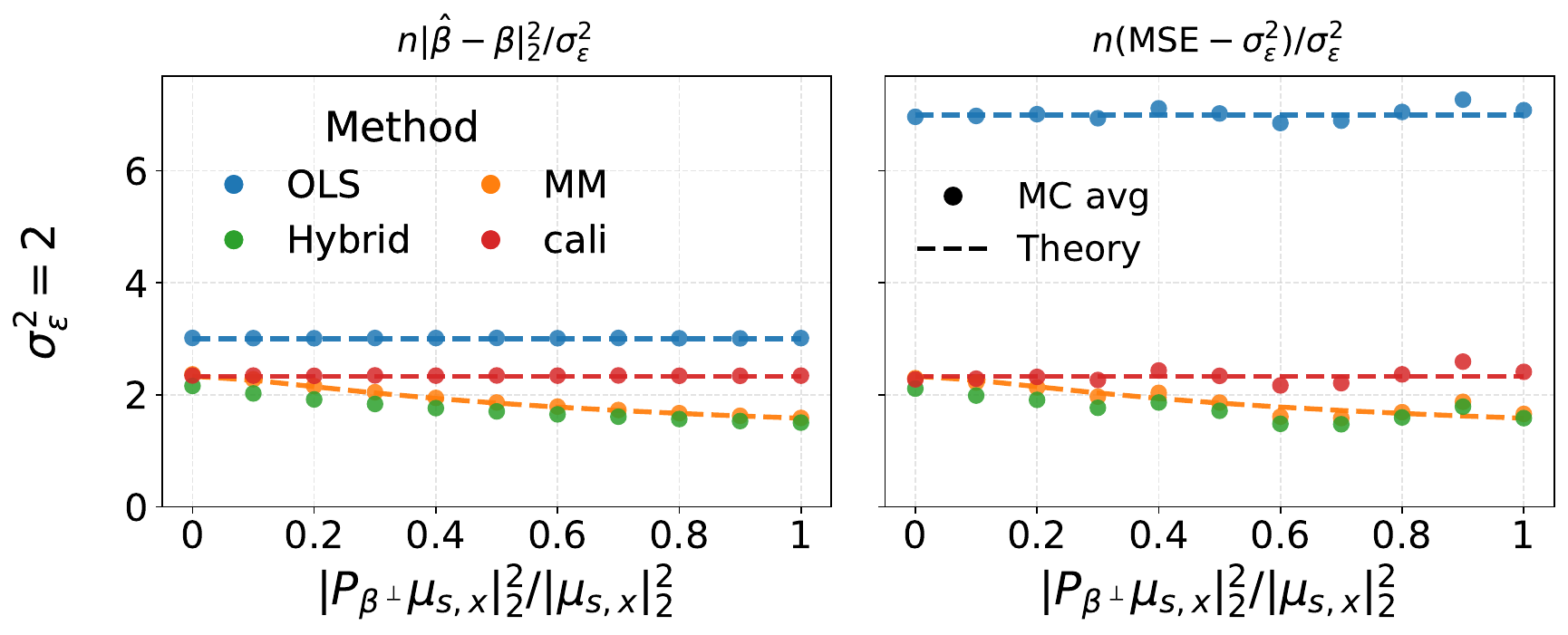}
    \caption{\textbf{Mean mismatch} ($\sigma_\varepsilon^2=2$). Scaled errors vs.\ $\rho=\|P_{\beta^\perp}\mu_{s,x}\|^2/\|\mu_{s,x}\|^2$. \rev{Monte Carlo standard errors are on the order of $10^{-3}$ and are visually negligible at this scale.}}
    \label{fig:toy_mu_sx}
\end{figure}

Figure~\ref{fig:toy_mu_sx} shows $\sigma_\varepsilon^2=2$: as more of the mean shift moves into the $\beta^\perp$ direction (larger $\rho$), moment-matching improves relative to two-stage, while the hybrid stays uniformly below both. OLS is flat at $4+\|\mu_{s,x}\|^2=7$; the target-aware methods are substantially lower. Across noise levels (Figure~\ref{fig:supp_mu_sx} in the Supplement), the two-stage MSE is flat in $\rho$ at every $\sigma_\varepsilon^2$ since it depends on $\Sigma_{s,x}$ but not on $\mu_{s,x}$.

\smallskip
\textbf{Accuracy-runtime tradeoffs.}
The theory predicts that two-stage and moment-matching achieve nearly the same accuracy as the hybrid at a fraction of the computational cost. To quantify this empirically, we run each estimator across noise levels $\sigma_\varepsilon^2\in\{1,\ldots,5\}$ and record wall-clock time per fit, varying the hybrid's convergence tolerance across three settings.

\begin{figure}[h]
    \centering
        \includegraphics[width=0.95\textwidth]{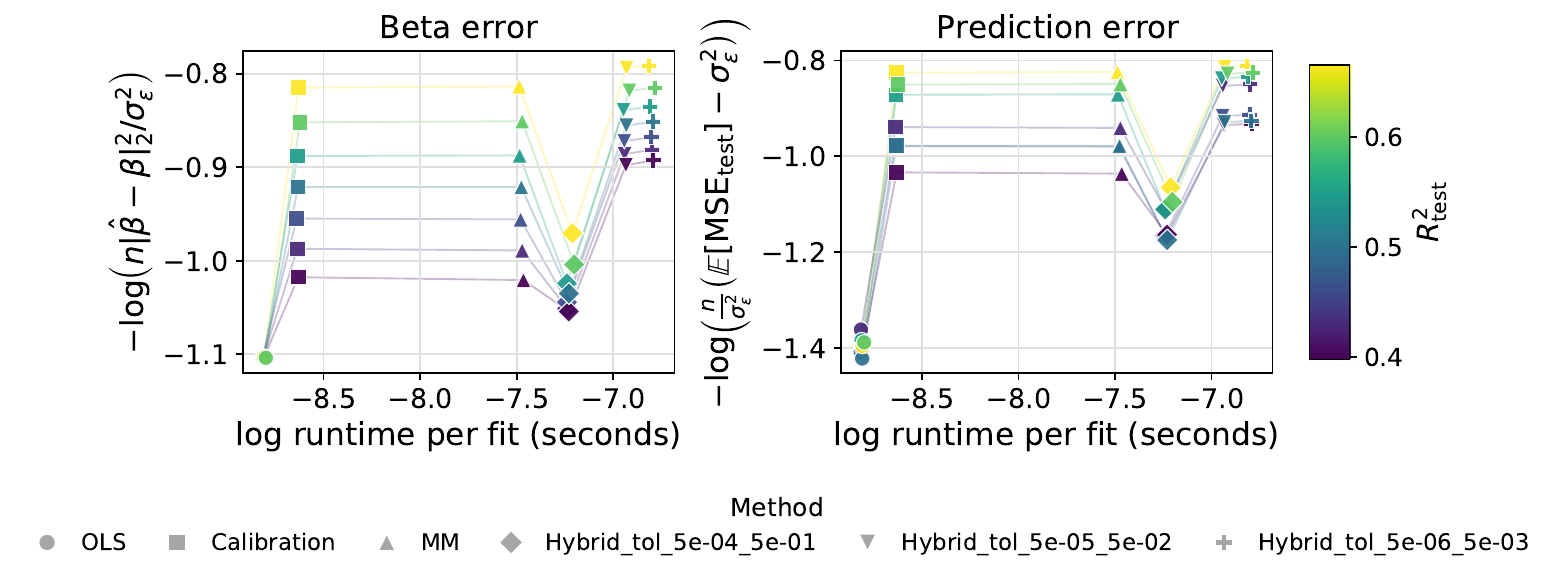}
    \caption{\textbf{Accuracy--runtime Pareto frontier.} Each point is one estimator at one noise level ($\sigma_\varepsilon^2\in\{1,\ldots,5\}$). Higher $y$-axis values indicate better accuracy; color indicates $R_{\mathrm{test}}^2$, which serves as a proxy for the signal-to-noise ratio. \rev{The three hybrid variants use the same sample plug-in $\omega^\star$ and are optimized by L-BFGS-B from the OLS initialization with maximum iteration count $1000$; their labels report the L-BFGS-B stopping tolerances $(\texttt{ftol},\texttt{gtol})$: $\texttt{Hybrid\_tol\_5e-04\_5e-01}$ uses $(5\times10^{-4},5\times10^{-1})$, $\texttt{Hybrid\_tol\_5e-05\_5e-02}$ uses $(5\times10^{-5},5\times10^{-2})$, and $\texttt{Hybrid\_tol\_5e-06\_5e-03}$ uses $(5\times10^{-6},5\times10^{-3})$.} 
    \rev{Monte Carlo standard errors are on the order of $10^{-3}$ and are visually negligible at this scale.}}
    \label{fig:pareto}
\end{figure}

Figure~\ref{fig:pareto} shows that OLS and the two-stage estimator lie at the far left of the curve (fastest), with the two-stage method attaining substantially higher accuracy at essentially the same computational cost. The hybrid estimator traces a Pareto frontier across tolerance levels, with some looser settings already capturing most of the attainable gains. Overall, among the fast methods, the two-stage estimator delivers striking accuracy per unit of computation at high signal-to-noise ratios.

\rev{\textbf{Computability.}}
\rev{All experiments were conducted on an Intel Core i7-14700KF (3.40GHz, 32GB RAM) and an NVIDIA GeForce RTX 4080 GPU (16GB).} Codes for reproducing results in this paper can be found at \url{https://github.com/TZstats-Columbia/target-aware-linear-regression}.

\section{Discussion}

We studied how target marginal information improves regression under distribution shift. Asymptotically, the hybrid estimator is uniformly best. \rev{Among the three target-aware estimators studied here, whether it achieves a semiparametric efficiency bound remains an open question.} Moment-matching and two-stage coincide in the isotropic case but diverge under joint mean and covariance shifts. Monte Carlo experiments corroborate these rankings across regimes.

Estimator choice depends critically on the signal-to-noise ratio.
\rev{The improvement conditions differ by estimator: the moment-matching estimator improves over OLS when $\sigma_\varepsilon^2 < 2\beta^\top Q_{s\mid k}^{-1}\beta$ (Corollary~\ref{cor:wdl_multi_beta_mse}), where $Q_{s\mid k}$ captures both the source covariance and the mean shift.}
\rev{The two-stage estimator improves when $\sigma_\varepsilon^2 < 2\beta^\top\Sigma_{s,x}^{-1}\beta$ (Theorem~\ref{thm:cali_multi}), which depends only on $\Sigma_{s,x}$.}
\rev{In the isotropic, zero-shift setting these conditions coincide at $\sigma_\varepsilon^2 < 2\|\beta\|^2$, but they diverge under covariance or mean shifts.}
In the high signal-to-noise regime, when a strong base model is available, the two-stage estimator nearly matches the hybrid at a fraction of the computational cost (Figure~\ref{fig:pareto}). This yields a simple rule: when the base predictor is accurate, post hoc rescaling captures most of the benefit from distributional information, and the additional hybrid optimization offers only small gains. As base models strengthen (through more data, richer features, or deeper architectures), the near-optimality region for two-stage expands, making cheap post hoc calibration increasingly appealing. When training-time optimization is affordable and signal-to-noise is moderate, the hybrid remains preferable.

\rev{The asymptotic formulas also reveal how the benefit of target marginals scales with the covariate dimension~$d$.}
\rev{For estimating the $d$-dimensional slope $\beta$, the scaled coefficient MSE is $d$ for OLS and $d-1+\sigma_\varepsilon^2/(2\|\beta\|^2)$ for moment-matching and two-stage in the isotropic, zero-shift setting.}
\rev{The absolute reduction is approximately one degree of freedom, independent of~$d$, so the relative improvement is $O(1/d)$ and all target-aware estimators converge to OLS as $d\to\infty$.}
\rev{This is because $\sigma_{k,y}^2$ provides a single scalar constraint on the $d$-dimensional~$\beta$.}
\rev{Across all simulation regimes, we also observe consistent reductions in the test mean squared error for predicting $Y$ when incorporating target marginals, in addition to the gains attributable solely to improved estimation of $\beta$.}


Although our derived asymptotic MSE results rely on linearity and Gaussian marginals, the general finding is expected to be applicable for nonlinear models, as we observed in machine learning applications \citep{hou2025calibrating}: post hoc calibration is most effective in high signal-to-noise regimes and captures most of the value of target marginals.
\rev{While the closed-form MSE expressions we derive rely on the Gaussian-linear setting, the key structural insights stem from the auxiliary-information mechanism rather than from any specific distributional assumptions. In particular, the dependence of performance gains on the signal-to-noise ratio, the near-optimality of inexpensive post hoc calibration when the base model is already strong, and the complementary effects of imposing constraints on the mean and variance are expected to extend to more complex and realistic modeling scenarios.}
\rev{\citet{hou2025calibrating} observe exactly these patterns in nonlinear neural network models calibrated via the kernelized Stein discrepancy, without any Gaussian or linearity assumption.}
\rev{The present paper provides the theoretical explanation for why that pattern occurs.}

\bibliography{staix_TAR}
\bibliographystyle{plainnat}

\newpage
\begin{center}
\Large\bfseries Supplement
\end{center}

\bigskip

\appendix
\renewcommand{\thesection}{S.\arabic{section}}
\renewcommand{\thesubsection}{S.\arabic{section}.\arabic{subsection}}
\renewcommand{\thetheorem}{S.\arabic{theorem}}
\renewcommand{\thelemma}{S.\arabic{lemma}}
\renewcommand{\theproposition}{S.\arabic{proposition}}
\renewcommand{\thecorollary}{S.\arabic{corollary}}
\renewcommand{\theequation}{S.\arabic{equation}}
\renewcommand{\thefigure}{S\arabic{figure}}
\setcounter{section}{0}
\setcounter{equation}{0}
\setcounter{figure}{0}
\setcounter{theorem}{0}
\setcounter{lemma}{0}
\setcounter{proposition}{0}
\setcounter{corollary}{0}
\setcounter{subsection}{0}

\setlength{\abovedisplayskip}{10pt plus 3pt minus 3pt}
\setlength{\belowdisplayskip}{10pt plus 3pt minus 3pt}
\setlength{\abovedisplayshortskip}{6pt plus 3pt}
\setlength{\belowdisplayshortskip}{6pt plus 3pt}

We use the notation from Section~3 throughout: 
\begin{eqnarray*}
    Q_s&=&\E(\tilde X\tilde X^\top),\\
    Q_{s\mid k}&=&\Sigma_{s,x}+(\mu_{s,x}-\mu_{k,x})(\mu_{s,x}-\mu_{k,x})^\top, \\
    v_{\sigma\beta}&=&\Sigma_{k,x}\beta,\kappa=v_{\sigma\beta}^\top Q_{s\mid k}^{-1}v_{\sigma\beta}, \\ 
    \chi&=&v_{\sigma\beta}^\top Q_{s\mid k}^{-1}\Sigma_{k,x}Q_{s\mid k}^{-1}v_{\sigma\beta}, \\
    \mbox{and } \tilde v_{\sigma\beta}&=&\tilde\Sigma_{k,x}\theta.
\end{eqnarray*}

\section{Detailed Estimator Derivations}
\label{app:derivations}

\subsection{OLS prediction error under the target distribution}

Under Assumptions~\ref{assump:multi_linear_model}--\ref{assump:multi_training_sample}, $\sqrt{n}(\hat\theta^{\mathrm{OLS}}-\theta)
\Rightarrow \mathcal{N}(0,\sigma_\varepsilon^2 Q_s^{-1})$, so
\[
\E\|\hat\beta^{\mathrm{OLS}}-\beta\|^2
=
\frac{\sigma_\varepsilon^2}{n}\operatorname{tr}(\Sigma_{s,x}^{-1})
+o(n^{-1}).
\]

If $(X,Y)$ is independent of the training sample with $X\sim\mathcal{N}(\mu_{k,x},\Sigma_{k,x})$ and $Y=\beta_0+X^\top\beta+\varepsilon$, then
\begin{align}
\E\!\left[\left(Y-\tilde{X}^\top\hat\theta^{\mathrm{OLS}}\right)^2\right]
&=
\E\!\left[\varepsilon^2\right]
+
\E\!\left[\left(\tilde X^\top(\theta-\hat\theta^{\mathrm{OLS}})\right)^2\right]
\notag\\[6pt]
&=
\sigma_\varepsilon^2
+
\operatorname{tr}\!\left(
\E[\tilde X\tilde X^\top]\cdot\E[(\hat\theta^{\mathrm{OLS}}-\theta)(\hat\theta^{\mathrm{OLS}}-\theta)^\top]
\right)
\notag\\[6pt]
&=
\sigma_\varepsilon^2
+
\frac{\sigma_\varepsilon^2}{n}
\operatorname{tr}\!\left(
(\tilde\Sigma_{k,x}+\tilde\mu_{k,x}\tilde\mu_{k,x}^\top)Q_s^{-1}
\right)
+o(n^{-1}).
\label{eq:supp_ols_pred}
\end{align}

\noindent
Under the normalization $\mu_{k,x}=0$, $\Sigma_{k,x}=I_d$, this gives
\[
\sigma_\varepsilon^2
+
\frac{\sigma_\varepsilon^2}{n}
\left[
1+\operatorname{tr}\!\left(\Sigma_{s,x}^{-1}\right)
+
\mu_{s,x}^\top \Sigma_{s,x}^{-1}\mu_{s,x}
\right]
+ o(n^{-1}),
\]
which recovers \eqref{eq:OLS_prediction_error} in the main text after writing 
\[
\Sigma_{s,x}^{-1}=\Sigma_{k,x}\Sigma_{s,x}^{-1}\] 
and \[
\mu_{s,x}^\top\Sigma_{s,x}^{-1}\mu_{s,x}=(\mu_{k,x}-\mu_{s,x})^\top\Sigma_{s,x}^{-1}(\mu_{k,x}-\mu_{s,x}).
\]

\subsection{MM: constraint structure and minimizer selection}
\label{app:wdl_detail}

The MM loss $L_{\mathrm{MM}}(\theta)$ defined in \eqref{eq:wdl_loss_multi} is a sum of two nonnegative squares. It attains its minimum value zero if and only if
\[
\tilde{\mu}_{k,x}^\top\theta=\mu_{k,y}
\qquad\text{and}\qquad
\theta^\top\tilde\Sigma_{k,x}\theta
+
\frac{1}{n}\|\boldsymbol Y-\boldsymbol{\tilde X}\theta\|_2^2
=
\sigma_{k,y}^2.
\]

The first equation (mean constraint) gives $\beta_0=\mu_{k,y}-\mu_{k,x}^\top\beta$. Substituting into the second equation yields a quadratic in~$\beta$.

Define the empirical moments
\[
\hat M_{xx}
:=
\frac{1}{n}\sum_{i=1}^n \check X_i\check X_i^\top,
\qquad
\hat m_{xy}
:=
\frac{1}{n}\sum_{i=1}^n \check X_i\check Y_i,
\qquad
\hat m_{yy}
:=
\frac{1}{n}\sum_{i=1}^n \check Y_i^2,
\]
where $\check X_i:=X_i-\mu_{k,x}$ and $\check Y_i:=Y_i-\mu_{k,y}$. Let
\[
A_n:=\Sigma_{k,x}+\hat M_{xx},
\qquad
\rev{c_n:=\hat m_{yy}-\sigma_{k,y}^2.}
\]

\begin{lemma}[Zero-loss characterization]
\label{lem:wdl_zero_loss}
The loss $L_{\mathrm{MM}}(\theta)$ attains its minimum value zero if and only if the quadratic
\begin{equation}
q_n(\beta)
:=
\beta^\top A_n\beta-2\hat m_{xy}^\top\beta+c_n
=0
\label{eq:supp_wdl_quadratic}
\end{equation}
has a real solution. In that case, $\hat\Theta^{\mathrm{MM}}=\{\theta(\beta): q_n(\beta)=0\}$ where $\theta(\beta)=(\mu_{k,y}-\mu_{k,x}^\top\beta,\;\beta^\top)^\top$.
\end{lemma}

Since $A_n$ is positive definite, we complete the square:
\begin{equation}
q_n(\beta)
=
(\beta-A_n^{-1}\hat m_{xy})^\top A_n(\beta-A_n^{-1}\hat m_{xy})
-
\Delta_n,
\label{eq:supp_wdl_completed}
\end{equation}
where $\Delta_n:=\hat m_{xy}^\top A_n^{-1}\hat m_{xy}-c_n$.

\medskip

\noindent\textbf{Case $\Delta_n=0$.}\;
The quadratic \eqref{eq:supp_wdl_quadratic} has the unique solution $\beta=A_n^{-1}\hat m_{xy}$.

\medskip

\noindent\textbf{Case $\Delta_n>0$.}\;
When $d>1$, equation \eqref{eq:supp_wdl_quadratic} has infinitely many solutions forming an ellipsoid. The MM estimator selects the one that minimizes the residual sum of squares. On the constraint set $q_n(\beta)=0$, we have
$\frac{1}{n}\sum_{i=1}^n (\check Y_i-\check X_i^\top\beta)^2
=
\sigma_{k,y}^2-\beta^\top\Sigma_{k,x}\beta$.
Therefore, \eqref{eq:wdl_loss_multi} is equivalent to
\[
\max_{\beta\in\mathbb{R}^d}\ \beta^\top\Sigma_{k,x}\beta
\quad\text{s.t.}\quad
q_n(\beta)=0.
\]

\begin{proposition}[KKT characterization of the MM minimizer]
\label{prop:wdl_kkt}
Let $\rho_n:=\lambda_{\max}(A_n^{-1/2}\Sigma_{k,x}A_n^{-1/2})$. If $\Delta_n>0$, then there exists a unique $\hat\lambda\ge \rho_n$ such that
\begin{equation}
(\hat\lambda A_n-\Sigma_{k,x})\,\hat\beta^{\mathrm{MM}}
=
\hat\lambda\,\hat m_{xy},
\qquad
q_n(\hat\beta^{\mathrm{MM}})=0.
\label{eq:supp_wdl_kkt}
\end{equation}
Moreover, if $\hat\lambda>\rho_n$, then $\hat\beta^{\mathrm{MM}}=\hat\lambda\,(\hat\lambda A_n-\Sigma_{k,x})^{-1}\hat m_{xy}$.
\end{proposition}

\noindent
The Lagrange multiplier $\hat\lambda$ is found by substituting $\hat\beta^{\mathrm{MM}}=\hat\lambda(\hat\lambda A_n-\Sigma_{k,x})^{-1}\hat m_{xy}$ into the constraint $q_n(\hat\beta^{\mathrm{MM}})=0$ and solving the resulting scalar equation numerically.

\medskip

\noindent\textbf{Case $\Delta_n<0$.}\;
The quadratic has no real solution, so $\min L_{\mathrm{MM}}>0$. 
Define
\[
\bar{\check X}
:=
\frac{1}{n}\sum_{i=1}^n \check X_i,
\qquad
\bar{\check Y}
:=
\frac{1}{n}\sum_{i=1}^n \check Y_i.
\]

\begin{lemma}
\label{lem:wdl_multi_negative_quartic}
Suppose $\Delta_n<0$. 
For $t\in\mathbb{R}$, define
\[
\beta(t)
:=
A_n^{-1}\left(\hat m_{xy}-t\bar{\check X}\right),
\qquad
\beta_0(t)
:=
\mu_{k,y}-\mu_{k,x}^\top\beta(t)+t,
\]
and
\[
u_n(t)
:=
\hat m_{yy}
-
\hat m_{xy}^\top A_n^{-1}\hat m_{xy}
+
2\left(
\hat m_{xy}^\top A_n^{-1}\bar{\check X}
-
\bar{\check Y}
\right)t
+
\left(
1-\bar{\check X}^\top A_n^{-1}\bar{\check X}
\right)t^2.
\]
Then every element of $\hat\Theta^{\mathrm{MM}}$ has the form
$\left(\beta_0(t),\beta(t)\right)$
for some real $t$ satisfying
\begin{equation}
\left(2t+u_n'(t)\right)^2 u_n(t)
-
\sigma_{k,y}^2\left(u_n'(t)\right)^2
=0.
\label{eq:wdl_multi_negative_quartic}
\end{equation}
\end{lemma}

With probability tending to one, $\Delta_n>0$ when $\beta\neq 0$, so this case is asymptotically negligible.

\subsection{Hybrid: gradient and optimal weight}
\label{app:hybrid_detail}

The hybrid loss $L_{\mathrm{H}}(\theta;\omega)$ in \eqref{eq:hybrid_loss_multi} does not admit a closed-form minimizer. Let $s(\theta):=n^{-1}\|\boldsymbol Y-\boldsymbol{\tilde X}\theta\|_2^2$ and $t(\theta):=\sqrt{\theta^\top\tilde\Sigma_{k,x}\theta+s(\theta)}$. A direct differentiation gives the gradient
\begin{equation}
\begin{aligned}
\nabla_\theta L_{\mathrm H}(\theta;\omega)
=&\;
-\frac{2}{n}\boldsymbol{\tilde X}^\top(\boldsymbol Y-\boldsymbol{\tilde X}\theta)
+
2\omega\left(\tilde{\mu}_{k,x}^\top\theta-\mu_{k,y}\right)\tilde{\mu}_{k,x}
\\[6pt]
&+
2\omega\,\frac{t(\theta)-\sigma_{k,y}}{t(\theta)}
\left(
\tilde\Sigma_{k,x}\theta
-
\frac{1}{n}\boldsymbol{\tilde X}^\top(\boldsymbol Y-\boldsymbol{\tilde X}\theta)
\right).
\end{aligned}
\label{eq:supp_hybrid_grad}
\end{equation}

\noindent
In practice, we solve $\nabla_\theta L_{\mathrm H}=0$ by gradient descent initialized at $\hat\theta^{\mathrm{OLS}}$.

\paragraph{Optimal $\omega^\star$ via quartic equation.}

By Corollary~\ref{cor:hybrid_multi_beta_mse}, the asymptotic MSE is $(\sigma_\varepsilon^2/n)\,V_{\mathrm H}(\omega)+o(n^{-1})$. To find the optimal weight, define
\[
m_1
:=
\tilde\mu_{k,x}^\top Q_s^{-1}\tilde\mu_{k,x},
\qquad
m_2
:=
\frac{1}{\sigma_{k,y}}
\tilde\mu_{k,x}^\top Q_s^{-1}\tilde v_{\sigma\beta},
\qquad
m_3
:=
\frac{1}{\sigma_{k,y}^2}
\tilde v_{\sigma\beta}^\top Q_s^{-1}\tilde v_{\sigma\beta},
\]
\[
n_1
:=
\tilde\mu_{k,x}^\top
Q_s^{-1}P_\beta^\top P_\beta Q_s^{-1}\tilde\mu_{k,x},
\qquad
n_2
:=
\frac{1}{\sigma_{k,y}}
\tilde\mu_{k,x}^\top
Q_s^{-1}P_\beta^\top P_\beta Q_s^{-1}\tilde v_{\sigma\beta},
\]
\[
n_3
:=
\frac{1}{\sigma_{k,y}^2}
\tilde v_{\sigma\beta}^\top
Q_s^{-1}P_\beta^\top P_\beta Q_s^{-1}\tilde v_{\sigma\beta},
\]
and the denominator polynomial
\[
\Delta_{\mathrm H}(\omega)
:=
1+(m_1+m_3)\omega+(m_1m_3-m_2^2)\omega^2.
\]

\begin{proposition}[Rational form of $V_{\mathrm H}$]
\label{prop:VH_rational}
Under Assumptions~\ref{assump:multi_linear_model}--\ref{assump:multi_knowledge_distribution},
\begin{equation}
V_{\mathrm H}(\omega)
=
\operatorname{tr}\!\left(P_\beta Q_s^{-1}P_\beta^\top\right)
+
\frac{
A_1\omega
+
A_2\omega^2
+
A_3\omega^3
+
A_4\omega^4
}{
\Delta_{\mathrm H}(\omega)^2
},
\label{eq:supp_VH_rational}
\end{equation}
where
\begin{align}
A_1
&=
-2(n_1+n_3),
\notag\\[6pt]
A_2
&=
-(n_1+n_3)(m_1+m_3)
-
3(m_1n_3+m_3n_1-2m_2n_2)
+
\frac{\sigma_\varepsilon^2}{2\sigma_{k,y}^2}\,n_3,
\notag\\[6pt]
A_3
&=
-2(m_1+m_3)(m_1n_3+m_3n_1-2m_2n_2)
+
\frac{\sigma_\varepsilon^2}{\sigma_{k,y}^2}(m_1n_3-m_2n_2),
\notag\\[6pt]
A_4
&=
-(m_1m_3-m_2^2)(m_1n_3+m_3n_1-2m_2n_2)
+
\frac{\sigma_\varepsilon^2}{2\sigma_{k,y}^2}
(m_1^2n_3-2m_1m_2n_2+m_2^2n_1).
\notag
\end{align}

Any minimizer $\omega^\star>0$ satisfies the quartic equation
\begin{equation}
B_0
+
B_1\omega^\star
+
B_2(\omega^\star)^2
+
B_3(\omega^\star)^3
+
B_4(\omega^\star)^4
=
0,
\label{eq:supp_quartic}
\end{equation}
where
\begin{align}
B_0
&=
A_1,
\qquad
B_1
=
2A_2-(m_1+m_3)A_1,
\notag\\[4pt]
B_2
&=
3A_3-3(m_1m_3-m_2^2)A_1,
\notag\\[4pt]
B_3
&=
4A_4
+
(m_1+m_3)A_3
-
2(m_1m_3-m_2^2)A_2,
\notag\\[4pt]
B_4
&=
2(m_1+m_3)A_4
-
(m_1m_3-m_2^2)A_3.
\notag
\end{align}

\noindent
In practice, one collects the nonnegative real roots of \eqref{eq:supp_quartic} and compares $V_{\mathrm H}$ at those roots and at $\omega=0$.
\end{proposition}

\subsection{MM prediction error}
\label{app:wdl_prediction}

If $(X,Y)$ is independent of the training sample with $X\sim\mathcal{N}(\mu_{k,x},\Sigma_{k,x})$ and $Y=\beta_0+X^\top\beta+\varepsilon$, then by the same argument as \eqref{eq:supp_ols_pred},
\begin{equation}
\E\!\left[\left(Y-\tilde{X}^\top\hat\theta^{\mathrm{MM}}\right)^2\right]
=
\sigma_\varepsilon^2
+
\frac{\sigma_\varepsilon^2}{n}
\operatorname{tr}\!\left(
(\tilde\Sigma_{k,x}+\tilde\mu_{k,x}\tilde\mu_{k,x}^\top)\,
J_{\mu_k}^\top Q_{s\mid k}^{-1}\Omega\, Q_{s\mid k}^{-1} J_{\mu_k}
\right)
+
o(n^{-1}).
\label{eq:supp_wdl_pred}
\end{equation}

\subsection{Hybrid prediction error}
\label{app:hybrid_prediction}

Similarly, the prediction error of the hybrid estimator under the target distribution is
\begin{equation}
\E\!\left[\left(Y-\tilde{X}^\top\hat\theta^{\mathrm{H}}(\omega)\right)^2\right]
=
\sigma_\varepsilon^2
+
\frac{\sigma_\varepsilon^2}{n}
\operatorname{tr}\!\left(
\left(
\tilde{\Sigma}_{k,x}
+
\tilde{\mu}_{k,x}\tilde{\mu}_{k,x}^\top
\right)
Q(\omega)^{-1}\Omega(\omega)Q(\omega)^{-1}
\right)
+
o(n^{-1}).
\label{eq:supp_hybrid_pred}
\end{equation}

\section{Proofs}
\label{app:proofs}

\subsection{Proof of Theorem~\ref{thm:hybrid_multi_asymp}}

The proof proceeds via a mean value expansion of the gradient of $L_{\mathrm H}$.

\medskip

\emph{Consistency.}\;
The population criterion $L_{\mathrm H,\infty}(\theta;\omega):=\E[(Y-\tilde X^\top\theta)^2]+\omega(\tilde\mu_{k,x}^\top\theta-\mu_{k,y})^2+\omega(\sqrt{\theta^\top\tilde\Sigma_{k,x}\theta+\E[(Y-\tilde X^\top\theta)^2]}-\sigma_{k,y})^2$ has $\theta$ as its unique minimizer because the first term is strictly convex and equals $\sigma_\varepsilon^2$ only at the true parameter. Since $L_{\mathrm H}(\cdot;\omega)$ converges uniformly on compacts to $L_{\mathrm H,\infty}(\cdot;\omega)$ by the law of large numbers, standard M-estimation arguments give $\hat\theta^{\mathrm H}(\omega)\overset{p}{\to}\theta$.

\medskip

\emph{Linearization.}\;
Let $\psi_n(\theta):=\frac{1}{2}\nabla_\theta L_{\mathrm H}(\theta;\omega)$.
Define
\[
U_n:=\frac{1}{\sqrt{n}}\sum_{i=1}^n \tilde X_i\varepsilon_i,
\qquad
W_n:=\frac{1}{\sqrt{n}}\sum_{i=1}^n (\varepsilon_i^2-\sigma_\varepsilon^2).
\]
Since $\tilde\mu_{k,x}^\top\theta=\mu_{k,y}$ and $t(\theta)^2=\theta^\top\tilde\Sigma_{k,x}\theta+\frac{1}{n}\sum\varepsilon_i^2=\sigma_{k,y}^2+\frac{1}{\sqrt{n}}W_n$, a Taylor expansion gives
\[
\sqrt{n}\,\psi_n(\theta)
=
-U_n + \frac{\omega}{2\sigma_{k,y}^2}W_n\,\tilde v_{\sigma\beta} + o_p(1).
\]

\noindent
By the multivariate central limit theorem, $(U_n,W_n)\Rightarrow\mathcal{N}(0,\mathrm{diag}(\sigma_\varepsilon^2 Q_s,\,2\sigma_\varepsilon^4))$, since $\E(\tilde X\varepsilon)=0$, $\E[\tilde X\varepsilon(\varepsilon^2-\sigma_\varepsilon^2)]=0$, and $\Var(\varepsilon^2-\sigma_\varepsilon^2)=2\sigma_\varepsilon^4$. Hence
$-\sqrt{n}\,\psi_n(\theta)\Rightarrow Z$ with $\Var(Z)=\sigma_\varepsilon^2\Omega(\omega)$.

\medskip

\emph{Hessian limit.}\;
Differentiating $\psi_n$ and evaluating at a consistent intermediate point $\tilde\theta_n$ on the line segment between $\theta$ and $\hat\theta^{\mathrm H}(\omega)$ yields $\nabla_\theta\psi_n(\tilde\theta_n)\overset{p}{\to}Q(\omega)$, which is positive definite.

\medskip

\emph{Conclusion.}\;
The first-order condition $\psi_n(\hat\theta^{\mathrm H}(\omega))=0$ and the mean value expansion give
\[
\sqrt{n}(\hat\theta^{\mathrm H}(\omega)-\theta)
=
-\nabla_\theta\psi_n(\tilde\theta_n)^{-1}\sqrt{n}\,\psi_n(\theta)
\Rightarrow
\mathcal{N}(0,\,\sigma_\varepsilon^2 Q(\omega)^{-1}\Omega(\omega)Q(\omega)^{-1}).
\]

\subsection{Proof of Theorem~\ref{thm:wdl_multi_asymp}}

Write $\check X_i:=X_i-\mu_{k,x}$, $\check Y_i:=Y_i-\mu_{k,y}$, and define $\hat M_{xx}:=\frac{1}{n}\sum_{i}\check X_i\check X_i^\top$, $\hat m_{xy}:=\frac{1}{n}\sum_i \check X_i\check Y_i$, $\hat m_{yy}:=\frac{1}{n}\sum_i \check Y_i^2$, $A_n:=\Sigma_{k,x}+\hat M_{xx}$, and $c_n:=\hat m_{yy}-\sigma_{k,y}^2+\beta^\top\Sigma_{k,x}\beta$, so that the MM variance constraint becomes $q_n(\beta):=\beta^\top A_n\beta-2\hat m_{xy}^\top\beta+c_n=0$.

\medskip

\emph{Consistency.}\;
Under the true parameter, the MM loss is exactly zero. The constraint set is nonempty with probability $1-O(n^{-1})$ (shown by verifying the discriminant $\Delta_n=v_{\sigma\beta}^\top A_n^{-1}v_{\sigma\beta}\cdot(\beta^\top A_n\beta-c_n)>0$ with high probability). The RSS-minimizing element converges to $\beta$ by the continuous mapping theorem.

\medskip

\emph{Linearization.}\;
Define $G_n(b,\lambda):=(\lambda(A_nb-\hat m_{xy})-\Sigma_{k,x}b,\;q_n(b))^\top$. The MM first-order conditions are $G_n(\hat\beta^{\mathrm{MM}},\hat\lambda)=0$ where $\hat\lambda$ is a Lagrange multiplier converging to $1$. A mean value expansion gives
\[
\sqrt{n}
\begin{pmatrix}
\hat\beta^{\mathrm{MM}}-\beta\\[4pt]
\hat\lambda-1
\end{pmatrix}
=
-D^{-1}\sqrt{n}\,G_n(\beta,1)+o_p(1),
\qquad
D:=
\begin{pmatrix}
Q_{s\mid k} & v_{\sigma\beta}\\[4pt]
2v_{\sigma\beta}^\top & 0
\end{pmatrix}.
\]

\medskip

\emph{Score evaluation.}\;
Substituting the true parameter,
\[
\sqrt{n}\,G_n(\beta,1)
=
\begin{pmatrix}
-U_n\\[4pt]
W_n
\end{pmatrix},
\qquad
U_n:=\frac{1}{\sqrt{n}}\sum_i \check X_i\varepsilon_i,
\quad
W_n:=\frac{1}{\sqrt{n}}\sum_i (\varepsilon_i^2-\sigma_\varepsilon^2),
\]
which is jointly Gaussian with $\Var(U_n)=\sigma_\varepsilon^2 Q_{s\mid k}$ and $\Var(W_n)=2\sigma_\varepsilon^4$ (independent).

\medskip

\emph{Solving.}\;
Block inversion of $D$ yields
\[
\sqrt{n}(\hat\beta^{\mathrm{MM}}-\beta)
\Rightarrow
Q_{s\mid k}^{-1}U - \frac{v_{\sigma\beta}^\top Q_{s\mid k}^{-1}U}{\kappa}\,Q_{s\mid k}^{-1}v_{\sigma\beta}
- \frac{W}{2\kappa}\,Q_{s\mid k}^{-1}v_{\sigma\beta},
\]
whose covariance is $\sigma_\varepsilon^2 Q_{s\mid k}^{-1}\Omega\, Q_{s\mid k}^{-1}$ with $\Omega=Q_{s\mid k}-\kappa^{-1}v_{\sigma\beta}v_{\sigma\beta}^\top+(\sigma_\varepsilon^2/2)\kappa^{-2}v_{\sigma\beta}v_{\sigma\beta}^\top$.
Since $\hat\beta_0^{\mathrm{MM}}-\beta_0=-\mu_{k,x}^\top(\hat\beta^{\mathrm{MM}}-\beta)$, the joint result follows.

\rev{\subsection{Proof of Theorem~\ref{thm:cali_multi}}

We work under the normalization $\mu_{k,x}=0$ and $\Sigma_{k,x}=I_d$.
Let $\delta_n:=\hat\beta^{\mathrm{OLS}}-\beta$, $\tau_n:=\hat\sigma_\varepsilon^2-\sigma_\varepsilon^2$, and $r^2:=\|\beta\|^2$.
Since $\beta\neq 0$ and $\sigma_{k,y}^2=r^2+\sigma_\varepsilon^2$,
\[
b
=
\left(
\frac{(r^2-\tau_n)_+}{\|\beta+\delta_n\|^2}
\right)^{1/2}.
\]
By standard OLS theory, $\delta_n=O_p(n^{-1/2})$ and $\tau_n=O_p(n^{-1/2})$. Hence the positive part is inactive with probability tending to one, and Taylor expansion gives
\[
b
=
1-\frac{\beta^\top\delta_n}{r^2}
-\frac{\tau_n}{2r^2}
+o_p(n^{-1/2}).
\]
Therefore
\[
\sqrt n\left(\hat\beta^{\mathrm{cali}}-\beta\right)
=
\sqrt n\delta_n
-
\frac{\beta^\top\sqrt n\delta_n}{r^2}\beta
-
\frac{\sqrt n\tau_n}{2r^2}\beta
+o_p(1).
\]

Conditional on the design, the OLS coefficients and the residual variance estimator are independent, and
$(n-d-1)\hat\sigma_\varepsilon^2/\sigma_\varepsilon^2\sim\chi^2_{n-d-1}$.
Thus
\[
\sqrt n\delta_n \Rightarrow Z,
\qquad
\sqrt n\tau_n \Rightarrow W,
\]
where $Z\sim\mathcal N(0,\sigma_\varepsilon^2\Sigma_{s,x}^{-1})$, $W\sim\mathcal N(0,2\sigma_\varepsilon^4)$, and $Z\perp W$.
It follows that
\[
\sqrt n\left(\hat\beta^{\mathrm{cali}}-\beta\right)
\Rightarrow
Z-\frac{\beta^\top Z}{r^2}\beta-\frac{W}{2r^2}\beta .
\]
The limiting second moment is
\[
\begin{aligned}
\E\left\|
Z-\frac{\beta^\top Z}{r^2}\beta-\frac{W}{2r^2}\beta
\right\|^2
&=
\E\|Z\|^2
-
\frac{\E(\beta^\top Z)^2}{r^2}
+
\frac{\E W^2}{4r^2}  \\
&=
\sigma_\varepsilon^2\operatorname{tr}\!\left(\Sigma_{s,x}^{-1}\right)
-
\sigma_\varepsilon^2
\frac{\beta^\top\Sigma_{s,x}^{-1}\beta}{\|\beta\|^2}
+
\frac{\sigma_\varepsilon^4}{2\|\beta\|^2}.
\end{aligned}
\]
The required uniform integrability follows from the sub-Gaussian assumption on $X$ and the Gaussianity of $\varepsilon$. Hence
\[
\E\!\left[\left\|\hat\beta^{\mathrm{cali}}-\beta\right\|^2\right]
=
\frac{\sigma_\varepsilon^2}{n}
\left[
\operatorname{tr}\!\left(\Sigma_{s,x}^{-1}\right)
-
\frac{\beta^\top\Sigma_{s,x}^{-1}\beta}{\|\beta\|^2}
+
\frac{\sigma_\varepsilon^2}{2\|\beta\|^2}
\right]
+
o(n^{-1}).
\]

For prediction, since the test covariates have empirical mean zero and empirical covariance $I_d$,
$\tilde Y_j=\mu_{k,y}+(X_j^{\mathrm{test}})^\top\hat\beta^{\mathrm{cali}}$.
Also $\mu_{k,y}=\beta_0$ under the normalization. Therefore
\[
\frac1m\sum_{j=1}^m
\E\!\left[(\tilde Y_j-Y_j^{\mathrm{test}})^2\right]
=
\sigma_\varepsilon^2
+
\E\!\left[\left\|\hat\beta^{\mathrm{cali}}-\beta\right\|^2\right],
\]
which gives \eqref{eq:cali_prediction}.}

\section{Additional Experimental Results}
\label{app:additional_experiments}

\subsection{Covariance geometry across noise levels}

\begin{figure}[h]
    \centering
    \includegraphics[width=\textwidth]{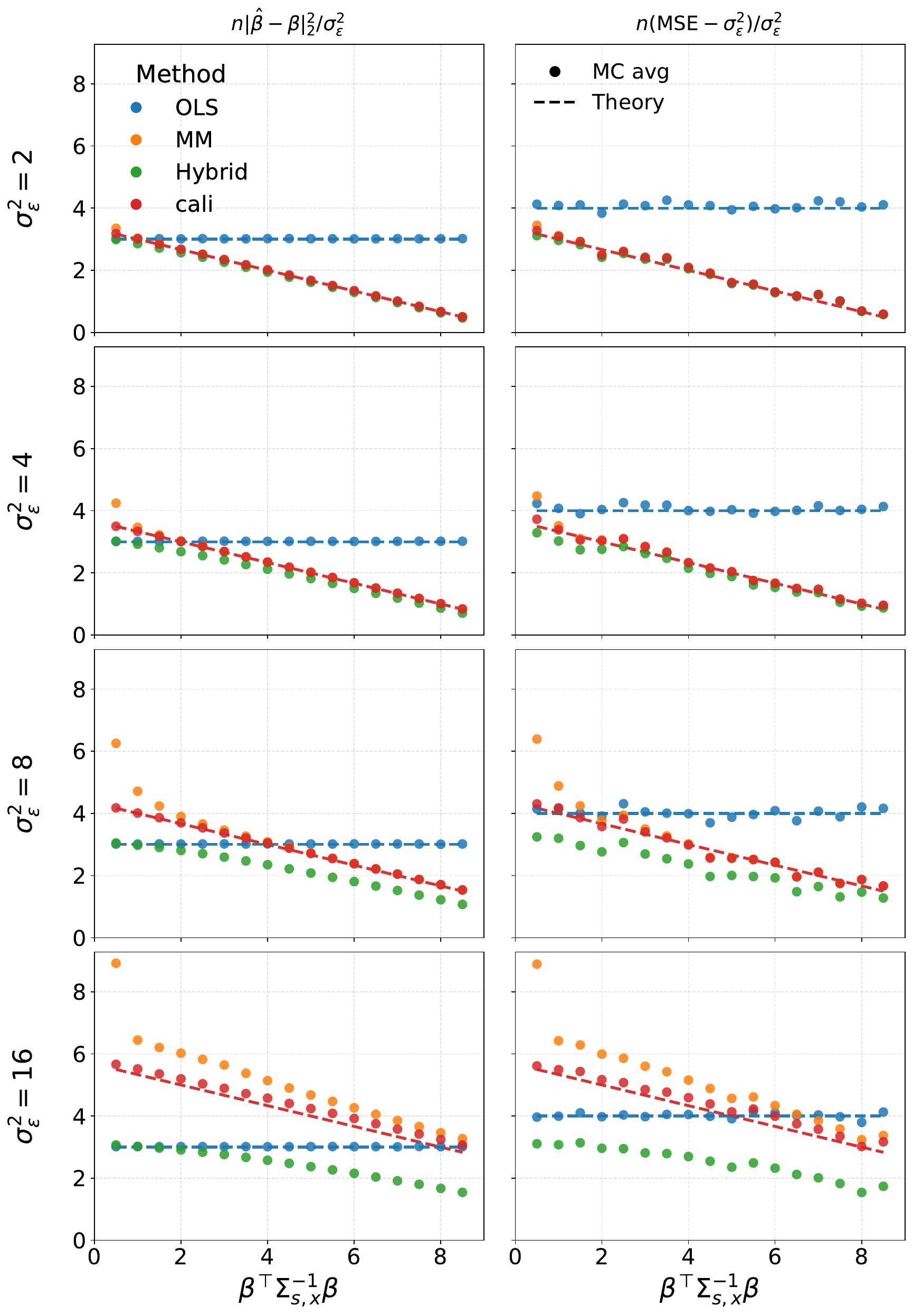}
    \caption{\textbf{Covariance geometry} across noise levels. Scaled coefficient error (left) and scaled excess prediction error (right) vs.\ $\beta^\top\Sigma_{s,x}^{-1}\beta$, for multiple values of $\sigma_\varepsilon^2$. Points: Monte Carlo averages ($L=10^6$, $n=1000$); dashed lines: theoretical values.}
    \label{fig:supp_sigma_sx}
\end{figure}

Figure~\ref{fig:supp_sigma_sx} displays the full covariance geometry experiment across multiple noise levels $\sigma_\varepsilon^2$. At low noise (high signal-to-noise ratio), the three target-aware estimators are nearly indistinguishable and all substantially improve over OLS. As the noise increases, the estimators separate: moment-matching (MM) continues to exploit the full geometry of $Q_{s\mid k}$ and improves steadily as $\beta^\top\Sigma_{s,x}^{-1}\beta$ grows, whereas the two-stage estimator, which projects only onto $\beta$ through $\Sigma_{s,x}$, improves more slowly. The hybrid remains uniformly best across all noise levels and geometry configurations.

\subsection{Mean mismatch across noise levels}

\begin{figure}[h]
    \centering
    \includegraphics[width=\textwidth]{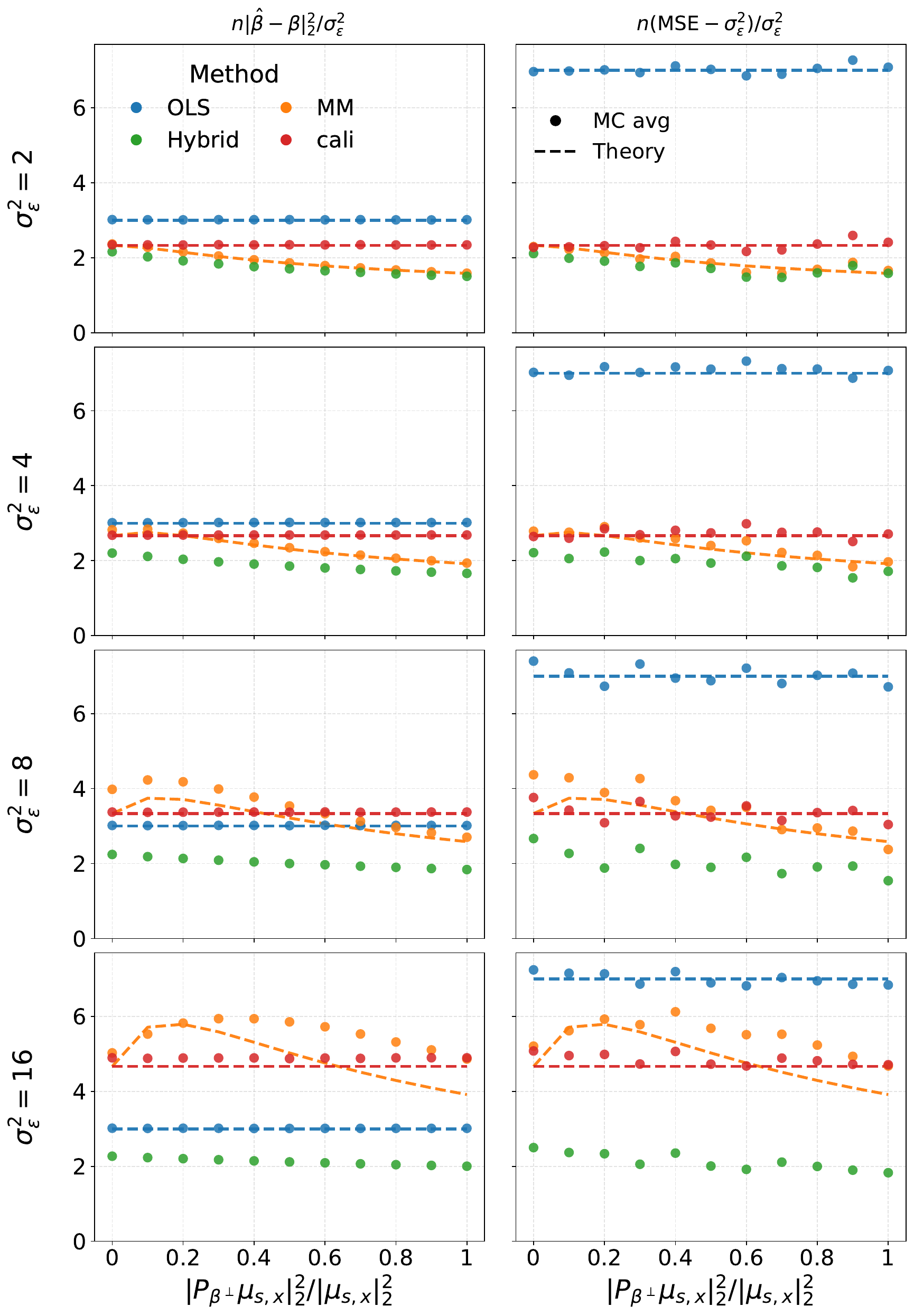}
    \caption{\textbf{Mean mismatch} across noise levels. Scaled errors vs.\ $\rho=\|P_{\beta^\perp}\mu_{s,x}\|^2/\|\mu_{s,x}\|^2$, for multiple values of $\sigma_\varepsilon^2$. Points: Monte Carlo averages ($L=10^6$, $n=1000$); dashed lines: theoretical values.}
    \label{fig:supp_mu_sx}
\end{figure}

Figure~\ref{fig:supp_mu_sx} displays the full mean mismatch experiment across multiple noise levels. The two-stage estimator is flat in $\rho$ at every noise level because its MSE \eqref{eq:cali_mse} depends on $\Sigma_{s,x}$ but not on $\mu_{s,x}$. Moment-matching, by contrast, benefits from mean shifts in the $\beta^\perp$ direction through $Q_{s\mid k}$, and this advantage grows with noise. At low noise all three target-aware estimators cluster together; at higher noise the ordering hybrid $<$ moment-matching $<$ two-stage becomes pronounced, with moment-matching gaining the most from favorable mean-shift geometry.

\end{document}